\documentclass{amsart}
\usepackage{graphicx}
\usepackage{natbib}
\usepackage[left,modulo]{lineno}
\theoremstyle{definition}

\theoremstyle{remark}

\numberwithin{equation}{section}

\begin{document}


\begin{linenumbers}

\title[Leaf litter decomposition]{Leaf litter decomposition -- Estimates of global variability based on Yasso07 model}%

\author[M. Tuomi et al.]{M. Tuomi$^{1,2}$}
\author[]{T. Thum$^{3}$}
\author[]{H. J\"arvinen$^{3}$}
\author[]{S. Fronzek$^{1}$}
\author[]{B. Berg$^{4,5}$}
\author[]{M. Harmon$^{6}$}
\author[]{J. A. Trofymow$^{7}$}
\author[]{S. Sevanto$^{8}$}
\author[]{J. Liski$^{1}$}

\thanks{{}\\
{1) Finnish Environment Institute, Research Programme for Global Change, Helsinki, Finland.}\\
{2) Department of Mathematics and Statistics, University of Helsinki, Helsinki, Finland.}\\
{3) Finnish Meteorological Institute, Helsinki, Finland.}\\
{4) Dipartimento Biologia Strutturale e Funzionale, Complesso Universitario, Monte San Angelo, Napoli, Italy.}\\
{5) Department of Forest Ecology, University of Helsinki, Helsinki, Finland.}\\
{6) Forest Ecology, Oregon State University, Corvallis, OR 97331, USA.}\\
{7) Canadian Forest Service, Pacific Forestry Centre, Victoria, British Columbia, Canada.}\\
{8) Department of Physics, University of Helsinki, Helsinki, Finland.}\\
{}\\
{}\\
{The corresponding author: M. Tuomi (mikko.tuomi@utu.fi; mikko.tuomi@ymparisto.fi)}\\
{Finnish Environment Institute, Research Programme for Global Change, Mechelininkatu 34a, P.O.Box 140, 00251 Helsinki, Finland}\\
{Phone: +358400148605, Fax: +358204902390}\\}

\keywords{Bayesian inference -- climate change -- decomposition -- leaf litter -- Markov chain Monte Carlo -- soil carbon}

\date{12.5.2009}%

\begin{abstract}
Litter decomposition is an important process in the global carbon cycle. It accounts for most of the heterotrophic soil respiration and results in formation of more stable soil organic carbon (SOC) which is the largest terrestrial carbon stock. Litter decomposition may induce remarkable feedbacks to climate change because it is a climate-dependent process. To investigate the global patterns of litter decomposition, we developed a description of this process and tested the validity of this description using a large set of foliar litter mass loss measurements (nearly 10 000 data points derived from approximately 70 000 litter bags). We applied the Markov chain Monte Carlo method to estimate uncertainty in the parameter values and results of our model called Yasso07. The model appeared globally applicable. It estimated the effects of litter type (plant species) and climate on mass loss with little systematic error over the first 10 decomposition years, using only initial litter chemistry, air temperature and precipitation as input variables. Illustrative of the global variability in litter mass loss rates, our example calculations showed that a typical conifer litter had 68\% of its initial mass still remaining after two decomposition years in tundra while a deciduous litter had only 15\% remaining in the tropics. Uncertainty in these estimates, a direct result of the uncertainty of the parameter values of the model, varied according to the distribution of the litter bag data among climate conditions and ranged from 2\% in tundra to 4\% in the tropics. This  reliability was adequate to use the model and distinguish the effects of even small differences in litter quality or climate conditions on litter decomposition as statistically significant.
\end{abstract}

\maketitle



\section{Introduction}

Litter decomposition plays a crucial role in the global carbon cycle. Carbon dioxide emissions from the decomposition of soil organic carbon (SOC) are equal to about 60 Pg of carbon per year, which is about seven times as much as the annual emissions of fossil carbon \citep{ipcc2007}. Most of these emissions originate from the decomposition of the relatively labile litter. Litter decomposition results also in formation of more stable organic compounds. These compounds represent the majority of all SOC \citep{davidson2006}, which is the largest terrestrial carbon stock, equal to about 2300 Pg \citep{jobbagy2000}, or three times the atmospheric carbon stock today \citep{ipcc2007}.

Litter decomposition will respond to changes in climate and, because of its great importance to the global carbon cycle, there may be remarkable feedbacks to the future climate change. Litter decomposition is also considered a complex process controlled by numerous other factors. To understand this process and to improve the estimates of the role of litter decomposition in the global carbon cycle, a global model of litter decomposition is needed.

A variety of approaches has been applied to estimate litter decomposition at large geographical scales. These approaches can be divided into three groups, 1) regression models based on litter bag studies \citep[e.g.][]{meentemeyer1978,berg1993,trofymow2002,zhang2007} or on soil respiration measurements \citep[e.g.][]{lloyd1994}, 2) specific dynamic soil carbon models comprising of compartments \citep[e.g.][]{parton1987,jenkinson1990} or models based on a theory of continuous SOC quality \citep{bosatta2003} and 3) less specific dynamic soil carbon models used for investigating the dynamics of nitrogen mineralization \citep{parton2007,manzoni2008} or applied to national carbon accounting \citep[e.g.][]{kurz2006,liski2005} or to dynamic global vegetation models \citep[e.g.][]{sitch2003} or Earth System Models (ESM) \citep[e.g.][]{jones2005}.

A particular challenge in developing SOC decomposition models is that the internal SOC pools of the models and especially carbon fluxes between the pools cannot easily be determined from measurement data \citep{christensen1996,elliott1996}. Therefore, it is a common practice to make a decision of the fluxes first and then quantify the parameters determining the magnitudes of these fluxes based on measurement data \citep[e.g.][]{moorhead1999}.

To investigate the global patterns of litter decomposition, we wanted to avoid this uncertainty stemming from the prefixed model fluxes and assumed that any fluxes between the SOC pools of our model were possible. Then, we determined these fluxes and the values of all other model parameters directly from a global data set of litter mass loss measurements \citep{gholz2000,trofymow1998,berg1991a,berg1991b,berg1993} using the Markov chain Monte Carlo (MCMC) method with Metropolis-Hastings algorithm \citep{metropolis1953,hastings1970}. We also required that the same parameter values should fit to the entire global data set so that our litter decomposition model would be applicable across climate conditions worldwide and to a wide variety of litter types.

Our aim was to calculate the full posterior probability density of model parameters, i.e. the joint probability density of all the free parameters in the model given the measurements. Possibilities to calculate such statistical uncertainty estimates have been lacking from previous SOC models although they are necessary if we want to evaluate the reliability of model results in a statistical sense. The uncertainty estimates can only be calculated in a meaningful way if the model is not over-parameterized. When building the model we required that there were no modelling errors introduced by over-parameterization. We also required that the model we developed fulfilled the Occamian principle of parsimony, which means that we chose the simplest model structure from a set of almost equally good alternatives. These requirements were made to find a reliable model whose prediction uncertainties are known reliably.

The objectives of this study were, first, to develop a global model of foliage litter decomposition that met the above requirements, and second, to estimate the global patterns of foliage litter decomposition and their uncertainty using this model.

\section{Material and methods}

\subsection{Measurements}

\begin{figure}
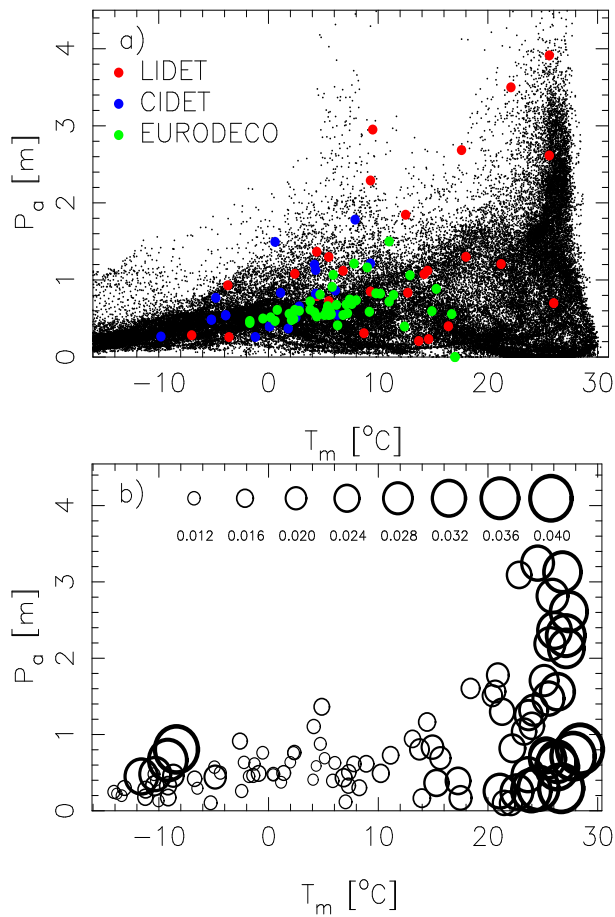

\centering
\includegraphics[angle=270, width=8.0cm, totalheight=6.0cm]{climatemap_tmpa.ps}

\includegraphics[angle=270, width=8.0cm, totalheight=6.0cm]{error_r3b.ps}
\caption{a) Distribution of the global climate conditions on land (black dots) from the CRU 2.1 database \citep{new2002} and the 97 litterbag measurement sites (LIDET sites red, CIDET sites blue and EURODECO sites green). Variables $T_{m}$ and $P_{a}$ are the mean annual temperature and annual precipitation, respectively. LIDET, CIDET and EURODECO measurement sites are in the USA and Central America, Canada, and Europe, respectively. b) 95\% confidence intervals as a function of climate for estimates of mass remaining after two years of decomposition (a unitary initial mass) calculated using the Yasso07 model. The error is shown for 100 randomly selected climatic conditions.}
\label{temperature_precipitation_scale}
\end{figure}

To develop the global model of foliage litter decomposition, we used litter bag data from 97 sites across Europe, and North and Central America (Table \ref{data2}). The data from the USA and Central America were collected within the LIDET network \citep{gholz2000} and the data from Canada within the CIDET network \citep{trofymow1998}, whereas the data from Europe originated from several research projects \citep{berg1991a,berg1991b,berg1993}. The study sites covered a wide range of climate conditions in terms of temperature and precipitation (Fig. \ref{temperature_precipitation_scale}a), which are the most important climate factors affecting litter decomposition \citep{meentemeyer1978,berg1993,aerts1997,liski2003,parton2007}.

The litter bag datasets consisted of measurements for foliage litter of 34 plant species including several coniferous and deciduous trees (Table \ref{data1}). The initial chemical composition was measured for each litter type and the loss of mass was followed for 3.1 to 10.2 years \citep{gholz2000,trofymow1998,berg1991a,berg1991b,berg1993}. In addition to the total mass loss, the mass loss of chemical compound groups was measured at seven Swedish study sites \citep{berg1991a,berg1991b}. Together the datasets comprised 9605 data points. The LIDET data consisted of values for individual litter bags, whereas the Canadian data points were averages of four litter bags collected at the same time \citep{trofymow2002}. Most of the European data points were averages of  25 litter bags \citep{berg1991a,berg1991b}. Thus, the data was received using approximately 70 000 litter bags.

In addition to the litter bag data, we used a dataset on accumulation of SOC at 26 sites along a 5500 year long soil chronosequence in southern Finland \citep{liski1998,liski2005}. This data provided us with both information on formation of humus from decomposing litter and humus decomposition.

\subsection{The model}

The Yasso07 model developed in this study is a generalization of an earlier Yasso soil carbon model \citep{liski2005}. Yasso07 is based on three assumptions of litter decomposition:
\begin{enumerate}
  \item Non-woody litter consists of four compound groups, i.e. compounds soluble in a non-polar solvent, ethanol or dichloromethane (denoted using E), or in water (W), and compounds hydrolysable in acid (A) and neither soluble nor hydrolyzable at all (N). Each group has its own mass loss rate independent of the origin of the litter. These compound groups are called the labile groups.
  \item The mass loss rates of the compound groups depend on the climatic conditions that can be described simply by using temperature and precipitation.
  \item Decomposition of the compound groups results in mass loss from the system and in mass flows between the compound groups. In addition, the mass loss of the four compound groups results in formation of more recalcitrant humus (H).
\end{enumerate}

The first assumption is justified by earlier studies showing that the above chemical groups decompose at different rates \citep{berg1982}, and that this grouping differentiates litter types according to decomposition rate \citep{palosuo2005}. The second assumption was based on results of several earlier studies \citep[e.g.][]{meentemeyer1978,berg1993,aerts1997,liski2003,parton2007}. The third assumption follows from a general view on carbon cycling in soil that involves transformations of organic compounds. This includes the break-down of complex compounds and the formation of simpler ones in chemical decomposition reactions, which are catalyzed by enzymes excreted by soil microbes; the uptake of the simplest compounds by soil microbes; and the formation of carbon dioxide and synthesis of biomass in the metabolism of soil microbes.

Yasso07 was formulated according to these hypotheses. All parameter values were treated as free parameters when the model was fitted to the data. Hence, Yasso07 is a set of first order differential equations and defined as
\begin{equation}\label{model DY}
    \dot{\mathbf{x}}(t) = \mathbf{A}(\mathbf{C}) \mathbf{x}(t) + \mathbf{b}(t) , \textrm{ } \mathbf{x}(0) = \mathbf{x}_{0}
\end{equation}
where $\mathbf{x} = (x_{A}, x_{W}, x_{E}, x_{N}, x_{H})^{T}$ is a vector describing the masses of the five compartments as a function of time ($t$); $\mathbf{A}(\mathbf{C})$ is a matrix describing the decomposition rates and the mass flows between the compartments as a function of climatic conditions ($\mathbf{C}$); vector $\mathbf{b}(t)$ is the litter input to the soil. Vector $\mathbf{x}_{0} = (x_{A,0},x_{W,0},x_{E,0},x_{N,0},x_{H,0})$ is the initial state of the system and $x_{i,0}$ the initial chemical composition, with $i$ referring to A, W, E, N and H. Matrix $\mathbf{A}$ is defined as a product of the mass flow matrix $\mathbf{A}_{p}$ and the diagonal decomposition coefficient matrix $\mathbf{k}(\mathbf{C}) = \textrm{diag} (k_{A},k_{W},k_{E},k_{N},k_{H})(\mathbf{C})$, where $k_{i}$ are the decomposition rate coefficients of the compartments. The mass flow matrix is defined as
\begin{equation}\label{model matrix}
    \mathbf{A}_{p} = \left( \begin{array}{ccccc}
     -1 & p_{1} & p_{2} & p_{3} & 0 \\
     p_{4} & -1 & p_{5} & p_{6} & 0 \\
     p_{7} & p_{8} & -1 & p_{9} & 0 \\
     p_{10} & p_{11} & p_{12} & -1 & 0 \\
     p_{H} & p_{H} & p_{H} & p_{H} & -1 \\
    \end{array} \right) ,
\end{equation}
where $p_{i} \in [0,1]$ are the relative mass flow parameters between the compartments. It is further assumed that $k_{i} = k_{i}(\mathbf{C})$, for all $i$, and that the sum of parameters $p_{i}$, describing the mass flows out of any of the compartments, does not exceed unity. This mass flow matrix was selected because all the possible flows between A, W, E and N are present. This is the most general model structure that can still be presented as a linear differential equation (Eq. \ref{model DY}).

The climate dependence of the decomposition rate factors $k_{i}$ is formulated as
\begin{equation}\label{climate dependence}
    k_{i}(\mathbf{C}) = \alpha_{i} \exp \big( \beta_{1}T + \beta_{2}T^{2} \big) \big( 1 - \exp [\gamma P_{a}] \big) ,
\end{equation}
where $T$ is temperature (Celcius scale) and $P_{a}$ is the annual precipitation and $\alpha_{i}$, $\beta_{1}$, $\beta_{2}$ and $\gamma$ are free parameters. This form of the temperature dependence was justified earlier by \citet{tuomi2008}.

We also tried to determine the separate values of these free parameters for each compound group, but it resulted in an over-parameterized model and a lower posterior probability. This happened because mass loss was measured by compound group only at the Swedish study sites (see Table \ref{data1}). Conseqeuently, we had to apply the same parameter values for each compound group.

The intra-annual variations in temperature resulting from seasonal changes were approximated by using a sinusoid
\begin{equation}\label{sinusoid T}
    T(t) = T_{m} + T_{a} \sin \Big( \frac{2 \pi t}{t_{P}} \Big),
\end{equation}
where $T_{a} = \frac{1}{2}( T_{m,max} - T_{m,min} )$ and $T_{m,max}$ and $T_{m,min}$ are the maximum and minimum mean monthly temperatures, respectively, and $t_{P}$ is the period of one year.

The full inverse solution, i.e. the probability density of the parameter vector $\theta$, was found by sampling the parameter space with the MCMC method. After a burn-in period, the proposal density was constructed to allow for the convergence of the chain to the posterior density with an acceptance rate of approximately 0.3. With the full probability density available, we then calculated the maximum \emph{a posteriori} (MAP) point estimate and Bayesian 95\% confidence set for the model parameter vector.

When calculating the posterior density, we used a Gaussian likelihood function for the measurements, such that
\begin{equation}\label{likelihood}
  f(\theta | m) \propto \exp \Bigg( -\frac{1}{2} \sum_{i} \omega_{i} S_{i} \Bigg) \pi(\theta) ,
\end{equation} 
where $f(\theta | m)$ is the conditional posterior density of the model parameters with the condition that measurements $m = (m_{1}, ..., m_{i}, ...)$ have been made and $\pi(\theta)$ is the prior density of the parameters. Variable $S_{i}$ is the common sum of squared residuals of measurement set $m_{i}$ and $\omega_{i}$ is some relative weight of measurements $m_{i}$. These weights were set to balance between data sets with large number of measurements containing little information and data sets with few measurements but high information content.

Finding the MAP estimate is equal to finding $\hat{\theta}$ that approximately satisfies the conditions
\begin{equation}\label{minimizationA}
    \hat{\theta} = \arg \min_{\theta} \sum_{i} \omega_{i} S_{i}
\end{equation}
and
\begin{equation}\label{minimizationB}
    \hat{\theta} = \arg \min_{\theta} S_{i}
\end{equation}
for all the data sets $m_{i}$ simultaneously. The weights $\omega_{i}$ were selected so that the latter condition is satisfied as well. This is basically made to extract all the information on model parameters from the measurements and to make sure that there is as little systematic error between the model and any of the measurement sets as possible.

We chose not to calculate the standard goodness-of-fit measures, the $r^{2}$ -values for the different datasets, because they provide no extra information in addition to the residuals between model with MAP parameter values and measurements. Also, to use the $r^{2}$ values, they should be compared with corresponding values calculated using different models. As there are no other models that predict the climate dependent time-evolution of the A, W, E, N and H compounds, this comparison could not be made.

We tested several alternative structures within the model framework set by our three assumptions, such as different precipitation dependence functions or different temperature or precipitation dependences for the different compound sets. These structures were compared using the model probabilities according to the Bayesian model selection theory, which automatically contains the Occam's razor and therefore penalises unnecessarily complicated model structures. These probabilities were defined as
\begin{equation}\label{model_probability}
  P(g_{j} | m) = \frac{P(m | g_{j}) P(g_{j})}{\sum_{k} P(m | g_{k}) P(g_{k})} , \textrm{  } P(m | g_{j}) = \int_{\theta_{j} \in \Omega_{j}} f(m | \theta_{j}, g_{j}) \pi(\theta_ {j} | g_{j}) d \theta_{j} ,
\end{equation}
where $g_{j}$ is the $j$th model structure, $\theta_{j} \in \Omega_{j}$ are its parameters, $P(g_{j} | m)$ is its likelihood and $P(g_{j})$ its prior probability. These prior probabilities were all set equal. We simply selected the structure that has the highest probability according to Eq. (\ref{model_probability}) with respect to all the measurements available. Despite that Bayesian methods have proven usefull in several statistical problems \citep[e.g.][]{ellison2004,lichter2005,tuomi2008}, according to our knowledge, this is the first time Bayesian methods are applied to the decomposition process of foliage litter.

When fitting the parameters of Yasso07 model to the data, we noticed that the level of mass loss rate was higher in the European litter bag measurements compared to the North or Central American ones. The same difference has been observed already earlier \citep{palosuo2005}. It is probably caused by a larger mesh size and smaller litter mass used in the European litter bags. This may have caused more leaching and thus overestimates of mass loss in Europe. To account for this difference, we introduced a scaling factor to the model and determined its value together with the other parameter values. This scaling factor had a narrow normal distribution with a mean equal to 0.58 and 95\% of the probability distribution between 0.56 and 0.60. The need to include the scaling factor in the analysis adds uncertainty to the results of this study. However, this uncertainty is related mostly to the estimated level of decomposition, whereas the estimated effects of climate and litter quality on decomposition are less dependent on the litter bag type. This happens because there were no systematic differences in the climate or litter quality effects between Europe and North and Central America despite the differences in the litter bags.

\section{Results}

We were able to establish a global description of foliage litter decomposition process using the large data set of litter bag measurements and advanced mathematical methods of complex inverse problems.

\begin{figure}
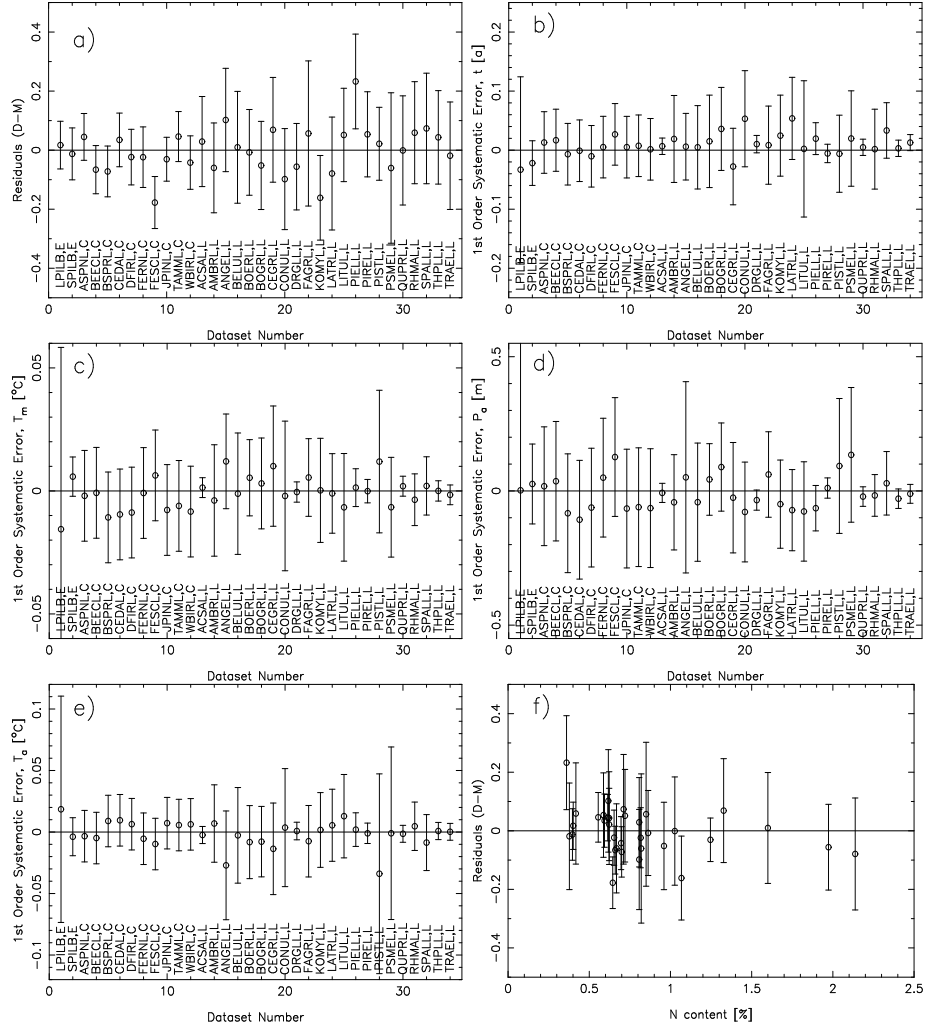

\centering
\includegraphics[angle=270, width=6.0cm, totalheight=4.5cm]{beta_h_l.ps}
\includegraphics[angle=270, width=6.0cm, totalheight=4.5cm]{beta_t_l.ps}

\includegraphics[angle=270, width=6.0cm, totalheight=4.5cm]{beta_e_l.ps}
\includegraphics[angle=270, width=6.0cm, totalheight=4.5cm]{beta_p_l.ps}

\includegraphics[angle=270, width=6.0cm, totalheight=4.5cm]{beta_s_l.ps}
\includegraphics[angle=270, width=6.0cm, totalheight=4.5cm]{beta_n_l.ps}
\caption{Model residuals (Data - Model) by litter type (a), linear 1st order systematic trends in the residuals by litter type as a function of time since the start of decomposition (b), annual mean temperature (c), annual precipitation (d), difference between the coldest and warmest months of the year (e), and the residuals as a function of the initial nitrogen concentration of the litter types (f). The error bars are 1$\sigma$ values, and thus they cover 68\% of the probability density. Letters E, C and L after dataset names refer to EURODECO, CIDET and LIDET, respectively.}
\label{residual_figures}
\end{figure}

The model fitted the data with little systematic error with respect to any variable investigated. The residuals, i.e. the differences between model-calculated estimates and measurements, did not deviate significantly from zero for 31 litter types out of 34 when looking at the 68\% confidence intervals (Fig. \ref{residual_figures}a). With a 95\% confidence level, none of the litter types deviated from zero (not shown) For one of the deviating types (\textit{Kobresia myosuroides}), there were only 20 data points, which makes it more probable that these measurements deviated by pure chance. The residuals were not correlated with either the time elapsed since the start of decomposition or any climate variable (Fig. \ref{residual_figures}b-e). The residuals were also uncorrelated with the initial nitrogen concentration of the litter species despite the fact that we did not include the effect of nitrogen in the model (Fig. \ref{residual_figures}f). The nitrogen concentration ranged from 0.3 to 2.2\%. The lack of systematic error means that the description of foliage litter decomposition in the model is, on average, valid for the global climate conditions and the range of litter characteristics covered by the litter bag data. The error in the results of the model is thus of random nature which can be characterized by probability densities with zero mean.

Fitting the model to the data revealed that there were four major carbon fluxes between the labile compartments (Table \ref{parameters}). Two of them were towards the compartments with higher decomposition rates representing breakdown of complex compounds and formation of simpler and labile ones during the decomposition process. The other two were towards the other direction indicating a re-synthesis of more complex compounds in the process. The other fluxes between the labile compartments were negligible. Four per cent of the combined mass loss of the labile compartments resulted in humus formation.

\begin{table}
\caption{Maximum \emph{a posteriori} parameter values of Yasso07 and the 95\% confidence limits.\label{parameters}}
\begin{tabular}{lccc}
  \hline \hline
    Parameter & Value & Unit & Interpretation \\
  \hline
    $\alpha_{A}$ & 0.66$\pm$0.11 & a$^{-1}$ & decomposition rate parameter of A \\
    $\alpha_{W}$ & 4.3$^{+1.6}_{-1.0}$ & a$^{-1}$ & decomposition rate parameter of W \\
    $\alpha_{E}$ & 0.35$\pm$0.08 & a$^{-1}$ & decomposition rate parameter of E \\
    $\alpha_{N}$ & 0.22$\pm$0.06 & a$^{-1}$ & decomposition rate parameter of N \\
    $p_{1}$ & 0.32$\pm$0.08 & - & relative mass flow magnitude, W $\rightarrow$ A\\
    $p_{2}$ & 0.01$^{+0.14}_{-0.01}$ & - & relative mass flow magnitude, E $\rightarrow$ A\\
    $p_{3}$ & 0.93$^{+0.03}_{-0.11}$ & - & relative mass flow magnitude, N $\rightarrow$ A\\
    $p_{4}$ & 0.34$^{+0.18}_{-0.15}$ & - & relative mass flow magnitude, A $\rightarrow$ W\\
    $p_{5}$ & 0.00$^{+0.07}_{-0.00}$ & - & relative mass flow magnitude, E $\rightarrow$ W\\
    $p_{6}$ & 0.00$^{+0.07}_{-0.00}$ & - & relative mass flow magnitude, N $\rightarrow$ W\\
    $p_{7}$ & 0.00$^{+0.01}_{-0.00}$ & - & relative mass flow magnitude, A $\rightarrow$ E\\
    $p_{8}$ & 0.00$^{+0.01}_{-0.00}$ & - & relative mass flow magnitude, W $\rightarrow$ E\\
    $p_{9}$ & 0.01$^{+0.07}_{-0.01}$ & - & relative mass flow magnitude, N $\rightarrow$ E\\
    $p_{10}$ & 0.00$^{+0.01}_{-0.00}$ & - & relative mass flow magnitude, A $\rightarrow$ N\\
    $p_{11}$ & 0.00$^{+0.06}_{-0.00}$ & - & relative mass flow magnitude, W $\rightarrow$ N\\
    $p_{12}$ & 0.92$^{+0.04}_{-0.15}$ & - & relative mass flow magnitude, E $\rightarrow$ N\\
    $\beta_{1}$ & 7.6$\pm$2.0 & $10^{-2}$ $^{\circ}$C$^{-1}$ & temperature dependence parameter \\
    $\beta_{2}$ & -8.9$\pm$6.5 & $10^{-4}$ $^{\circ}$C$^{-2}$ & temperature dependence parameter \\
    $\gamma$ & -1.27$\pm$0.20 & m$^{-1}$ & precipitation dependence parameter \\
    $p_{H}$ & 0.040$\pm$0.009 & - & mass flow to humus \\
    $\alpha_{H}$ & 3.3$^{+0.6}_{-0.7}$ & $10^{-3}$ a$^{-1}$ & humus decomposition coefficient \\
  \hline \hline
\end{tabular}
\end{table}

The decomposition rates were significantly different between the five compound groups included in the model (Table \ref{parameters}). The decomposition rate of the water soluble compounds was an order of magnitude higher than the corresponding rates of the other three labile compound groups. Among these groups, the decomposition rates decreased from the acid hydrolyzable compounds to the non-polar extractable and  the acid unhydrolyzable compounds. The decomposition rate of the humus compartment was two orders of magnitude lower.

\begin{figure}
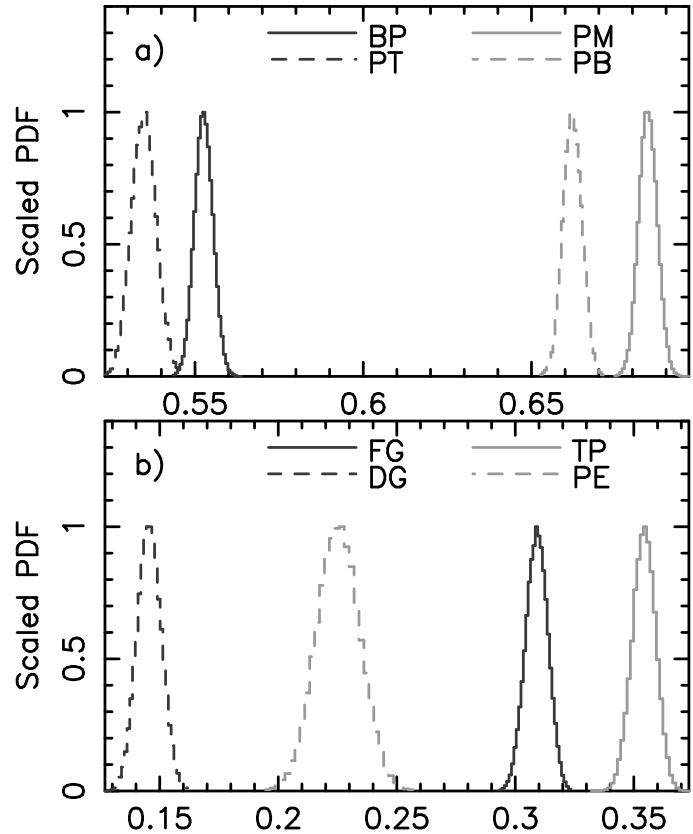

\centering
\includegraphics[angle=270, width=9.0cm]{cmr_nor_2YY_a0MT_harmaa.ps}

\includegraphics[angle=270, width=9.0cm]{cmr_sou_2YY_a0MT_harmaa.ps}
\caption{Probability densities of Yasso07 estimates for litter mass-remaining after two years of decomposition (a unitary initial mass) in different climate zones. A characteristic coniferous and deciduous species selected for each climate zone, tundra \textit{Picea mariana} (PM) and \textit{Betula papyrifera} (BP), the boreal zone \textit{Pinus banksiana} (PB) and \textit{Populus tremuloides} (PT), the temperate zone \textit{Thuja plicata} (TP) and \textit{Fagus grandifolia} (FG), and the tropics \textit{Pinus elliottii} (PE) and \textit{Drypetes glauca} (DG). The latitude, longitude, $T_{m}$, $T_{a}$ and $P_{a}$ of these sites are: tundra 56.32$^{\circ}$, -94.85$^{\circ}$, -4.0$^{\circ}$C, 10.1$^{\circ}$C, 0.50m; the boreal zone 53.22$^{\circ}$, -105.97$^{\circ}$, -2.4$^{\circ}$C, 9.7$^{\circ}$C, 0.59m; the temperate zone 35.00$^{\circ}$, -83.50$^{\circ}$ 13.9$^{\circ}$C 5.1$^{\circ}$C, 1.72m; the tropics 9.17$^{\circ}$, -79.85$^{\circ}$, 26.0$^{\circ}$C, 0.6$^{\circ}$C, 2.86m.}
\label{density_image}
\end{figure}

The calculated estimates for the mass-loss rates of foliage litter were statistically significantly different among sites selected to represent tundra, the boreal zone, the temperate zone and the tropics and, within each site, among a typical coniferous and deciduous litter types (Fig. \ref{density_image}). The probability densities of these estimates represent uncertainty caused by uncertainty about the parameter values of the model. The uncertainties of model predictions in Fig. \ref{temperature_precipitation_scale}b show the magnitude of 95\% confidence intervals as a function of climatic conditions for a decomposing body with an unitary initial mass. This uncertainty followed the distribution of the litter bag data across the global climate conditions (Fig. \ref{temperature_precipitation_scale}). It was smallest for the boreal and temperate zones where most of the study sites were located, and largest for the coldest tundra and the tropics which had only a few study sites.

\begin{figure}
\centering
\includegraphics[angle=270, width=12.0cm, totalheight=7.0cm]{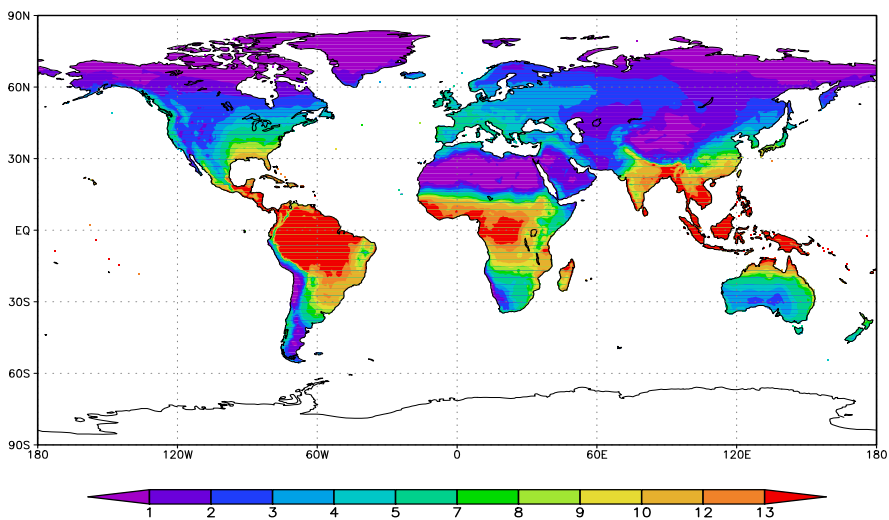}
\caption{Estimated geographic variations in the leaf litter decomposition rate. A unitary rate corresponds to climatic conditions where $T_{m} = 0^{\circ}$C, $T_{a} = 0^{\circ}$C and $P_{a}$ is sufficiently large to not limit the decomposition.}
\label{colour_maps}
\end{figure}

Illustrative of the global variability in the mass-loss rate of foliage litter, the coniferous litter of tundra (\emph{Picea mariana} needles) had about 68\% of its mass still remaining after two years of decomposition, while the deciduous litter of the tropics (\emph{Drypetes glauca} leaves) had only about 15\% remaining (Fig. \ref{density_image}). The values of the decomposition rate parameters of the model were up to 13 times higher in the tropics compared to the northern boreal zone (Fig. \ref{colour_maps}). The decomposition rates are not directly comparable to the mass loss rates because decomposition results also in transfer of carbon between the compartments inside the model.

\section{Discussion and conclusions}

The aim of this study was to estimate the global patterns of foliage litter decomposition after developing an appropriate model for this purpose. We required that the model was globally applicable and the results were accompanied by uncertainty estimates.

We found that the results of the Yasso07 model did not deviate systematically from the mass loss measurements of different litter types across the global climate conditions (Fig. \ref{residual_figures}) for 31 litter types out of 34 when looking at the 68\% confidence intervals (1$\sigma$) and for none of the litter types when looking at the 95\% intervals (2$\sigma$). The lack of the systematic deviation means that the probability densities of the results of the model represent the random uncertainty caused by uncertainty about the parameter values of the model (Fig. \ref{temperature_precipitation_scale}b, \ref{density_image}). Based on these results, we conclude that Yasso07 met the requirements we set for the global model.

We formulated the Yasso07 model according to three assumptions of foliage litter decomposition (see The Model). Comparison of the model results to the measurements provides us with means to evaluate the validity of these assumptions.

Regarding the first assumption, the decomposition rates of the four labile compound groups differed statistically significantly from each other (Table \ref{parameters}). When applying the same group-specific rates independent of litter type, the Yasso07 model estimated the mass loss rate of 31 litter types out of 34 without a statistically significant systematic error (Fig. \ref{residual_figures}a). These results support the validity of the first assumption. In addition, they suggest that the mass loss of different litter types can be estimated based on the initial division of the mass to the four compound groups over, at least, the first 10 years of decomposition, which was the time period covered by our data.

Regarding the second assumption, temperature and precipitation appeared indeed as the major climate factors affecting foliage litter decomposition at the global scale. After their effects were accounted for, the results of the Yasso07 model did not deviate from the mass loss measurements in any climate-related way (Fig. \ref{residual_figures}b to e). Temperature and precipitation, combined in various ways, have been used to explain the effects of climate on litter decomposition already earlier \citep{meentemeyer1978,berg1993,aerts1997,liski2003,parton2007}. We tested also different combinations of these two variables to find the final and best model. The Gaussian formulation of the temperature effects was based on our earlier comparison of alternative formulations \citep{tuomi2008}. Unfortunately, the data did not allow us to test whether the effects of temperature or precipitation differed between the compartments. Trying to fit separate values of these parameters resulted in an over-parameterized model and a lower posterior probability.

Regarding the third assumption, the information in the measurement, based on the parameter probability densities, was sufficient for distinguishing four major carbon fluxes between the labile compartments and a flux to the humus compartment in addition to the fluxes out of the system. These fluxes were identified without direct measurements of the fluxes themselves. Yet, our model fits to a general view of microbial litter decomposition and humus formation \citep{stevenson1982}. Two of the internal fluxes were towards more labile compartments representing probably a break-down of complex compounds by exoenzymes exudated by soil microbes. The flux from the water soluble materials to  acid hydrolysable ones may represent synthesis of more complex compounds in the metabolism of the soil microbes. The flux from the ethanol soluble compartment to the non-soluble one is more difficult to explain. It may be a result of the fitting procedure of the model to the data that led to rather similar decomposition rates of these compartments. Recalcitrant humus is formed from decomposition products in chemical reactions \citep{stevenson1982} explaining probably the flux to the humus compartment.

Process-modellers may consider it as a weakness that we applied the strongly empirical method without \emph{a priori} information to determine the fluxes of the model. However, it was the only method that we could use because direct measurements of the fluxes are lacking, especially at this extensive geographical scale. \citet{adair2008} ignored fluxes between carbon pools in their model when studying litter decomposition across North and Central America. However, the common view of the decomposition process \citep{stevenson1982} and more detailed decomposition models indicate that such internal fluxes exist \citep{parton1987}.

Determining fluxes between compartments is a common problem in developing dynamic compartmental soil carbon models because these fluxes cannot usually be measured \citep[e.g.][]{christensen1996,elliott1996}. For this reason, it is common to decide about the fluxes before determining the parameter values of the model \citep[e.g.][]{moorhead1999}. This practise adds uncertainty to the structure of the model that is difficult to control. To avoid this problem, we determined each flux of the Yasso07 model from the measurements (Table \ref{data1}). The ability to indentify the fluxes from the rather unspecific litter bag measurements is an indication of the power of the mathematical methods we used.

It was surprising that the Yasso07 model fitted to the data as well as it did even though the nitrogen concentration of the litter types or other characteristics of nitrogen availability at the sites, such as site fertility, were not accounted for in the model at all. Moreover, the residuals were not correlated with the initial nitrogen concentration of the litter species (Fig. \ref{residual_figures}f). Nitrogen is considered as an important regulator of decomposition and in many other models it is used as a controlling input variable \citep[e.g.][]{parton1987,jenkinson1990}. According to the results of this study, climate and chemical litter quality are still stronger controls of decomposition at the global scale. Once their effects were modelled carefully, the effects of nitrogen could not be distinguished from the remaining variability in the data. Nitrogen effects may still be important at smaller scales or in more detailed experiments.

It is difficult to compare the reliability of litter decomposition estimates calculated using Yasso07 and other models. The reason is that error in the results of the other models has not been assessed in a similar probabilistic sense. The regression models of litter decomposition have been evaluated mainly on the basis of Pearson correlation values or $r^2$ values \citep[e.g.][]{berg1993,aerts1997}, and little attention has been paid to the reliability of the results \citep[e.g.][]{tuomi2008}. Even if some $r^2$ values were available for other globally applicable models fitted to similar measurements, they could not be used for comparison easily. This is because the $r^2$ values depend on error estimates of individual measurements and if these errors have not been measured they need to be approximated or estimated by the modeller. More importantly, however, the $r^{2}$ values tell us nothing about the possible over-parameterization of the model. Increasing the number of free parameters in a model always improves the fit to the data and thus increases the $r^2$ value. However, errors introduced in the model structure at the same time may actually make the results of the model less reliable. For this reason, to assess the reliability of a model, it is necessary to ensure that the model is not over-parameterized. For this purpose, the Bayesian model probabilities provide a superior tool. To our knowledge, the only model that has been proven parsimonious in this sense, and is capable of describing litter decomposition process globally and presenting the results as statistical probability densities, is the one presented here.

Yasso07 is based on a biological process-oriented view of litter decomposition process although we used advanced mathematical methods to develop the model. A fundamental prerequisite we set to the model, was that its structure was not over-parametrized. This was necessary in order to determine unequivocal values for the parameters of the model. Such values were needed to obtain an unequivocal description of the biological process, test this description in the light of the litter bag data and calculate reliability estimates (probability densities) for the results of the model. This modelling approach we used is different from most of the earlier approaches taken to model litter decomposition at extensive geographical scales \citep[e.g.][]{berg1993,moorhead1999,trofymow2002,liski2003,adair2008}. However, it was the approach that made it possible for us to meet the objectives of this study. Alternatively, had we chosen the so-called mechanistical modelling approach, it would have been impossible to calculate uncertainty estimates for model predictions. The reason is that only some of the relevant processes can be measured directly and these measurements are only rarely available in global scale.

The global patterns of foliage litter decomposition calculated with this model, agreed broadly with earlier estimates \citep{matthews1997}. A detailed comparison was impossible, because the parameters represented in the studies were somewhat different, soil respiration \citep{matthews1997} vs. relative effect of climate on litter decomposition rate (Fig. \ref{colour_maps}). The global patterns were also quite similar to those of net primary productivity \citep{schuur2003,cramer1999}. This was expected because both processes depend strongly on temperature and precipitation.

The results of Yasso07 were reliable enough to distinguish the differences in litter mass loss rates between the sites representing four climate zones (tundra, boreal zone, temperate zone and tropics) and two litter types (coniferous, deciduous) inside each site. As a matter of fact, already a 3\% difference in remainig mass values (relative to the initial mass) appeared as statistically significant between the tundra and the boreal site. This difference was caused by only a 1.6 °C higher annual mean temperature at the boreal site. Similar climate warming is expected to take place as soon as over the next few decades at the high latitudes, and some regions have warmed by this much or more already \citep{ipcc2007}. Based on these differences in the decomposition rates between the current litter types and the currently prevailing climate conditions, Yasso07 model is reliable enough for distinguishing small changes in the litter decomposition rates over relatively short periods of time in response to climate change.

Using a large dataset of litter bag measurements and advanced mathematical methods of complex inverse problems we were able to develop a globally applicable description of foliage litter decomposition process and present this description in an exact mathematical form. This model, Yasso07, can be used for estimating the effects of changing climate and litter type on soil carbon decomposition in statistical sense. Our study supported the results of previous studies in that temperature and precipitation are the main factors affecting litter decomposition. Our results suggest, however, that nitrogen content of litter or site fertility parameters have a negligible effect on litter decomposition rates at the global scale. Because of the uncertainty estimates we produced for the parameter values of the model, Yasso 07 is currently the only model that can be used for evaluating how large the changes in the climate and/or litter type have to be before significant statistical differences can be detected by modelling.

We conclude that Yasso07 is suitable for estimating foliage litter decomposition of a wide variety of plant species across the global climate conditions. As a result of its general yet parsimonious structure, it can be used as a litter decomposition module in larger ecosystem models or Earth System Models.

\section*{Acknowledgements}

This study was initiated at a meeting of NCEAS Project 6840 “Analysis of long-term litter decomposition experiments: Synthesis at the site, regional, and global levels”. This study was funded by the Maj and Tor Nessling Foundation (project "Soil carbon in Earth System Models") and the Academy of Finland (project 107253). CIDET was funded through the Natural Resources Canada PERD Program and Canadian Forest Service Climate Change Network and Forest Carbon Project. M. Tuomi would like to acknowledge S. Kotiranta for help in editing Fig. 3.


\linespread{1.0}
\normalfont

\clearpage\newpage

\begin{table}
\caption{The litter bag study sites.\label{data2}}
\resizebox{\textwidth}{!}{%
\begin{tabular}{llcclllccl}
  \hline \hline
    Dataseries & Latitude & Longitude & Elevation & Ecosystem & Dataseries & Latitude & Longitude & Elevation & Ecosystem \\
    & [$^{\circ}$] & [$^{\circ}$] & [m] & & & [$^{\circ}$] & [$^{\circ}$] & [m] & \\
  \hline
LIDET	&	44.23	&	-122.18	&	500	&	Temperate Coniferous	&	EURODECO	&	60.63	&	13.62	&	400	&	Boreal	\\
LIDET	&	68.63	&	-149.57	&	760	&	Alpine Grassland	&	EURODECO	&	66.37	&	20.03	&	405	&	Boreal	\\
LIDET	&	9.17	&	-79.85	&	30	&	Wet Tropical	&	EURODECO	&	48.28	&	2.68	&	83	&	Temperate	\\
LIDET	&	64.75	&	-148.00	&	300	&	Boreal Coniferous	&	EURODECO	&	59.63	&	14.97	&	178	&	Boreal	\\
LIDET	&	38.87	&	-105.65	&	1300	&	Temperate Coniferous	&	EURODECO	&	43.67	&	7.97	&	447	&	Mediterranean	\\
LIDET	&	45.40	&	-93.20	&	230	&	Humid Grassland	&	EURODECO	&	52.68	&	23.78	&	165	&	Temperate	\\
LIDET	&	40.82	&	-104.77	&	1650	&	Dry Grassland	&	EURODECO	&	59.12	&	15.73	&	70	&	Boreal	\\
LIDET	&	35.00	&	-83.50	&	700	&	Temperate Deciduous	&	EURODECO	&	38.12	&	-6.20	&	2	&	Mediterranean	\\
LIDET	&	17.95	&	-65.87	&	80	&	Dry Tropical	&	EURODECO	&	64.30	&	16.33	&	330	&	Boreal	\\
LIDET	&	43.93	&	-71.75	&	300	&	Temperate Deciduous	&	EURODECO	&	52.03	&	5.70	&	45	&	Temperate	\\
LIDET	&	42.53	&	-72.17	&	335	&	Temperate Deciduous	&	EURODECO	&	53.00	&	9.95	&	81	&	Temperate	\\
LIDET	&	32.50	&	-106.75	&	1410	&	Shrubland	&	EURODECO	&	41.78	&	-5.43	&	760	&	Mediterranean	\\
LIDET	&	58.00	&	-134.00	&	100	&	Boreal Coniferous	&	EURODECO	&	56.43	&	14.58	&	140	&	Boreal	\\
LIDET	&	42.40	&	-85.40	&	288	&	Agriculture	&	EURODECO	&	39.42	&	9.25	&	80	&	Mediterranean	\\
LIDET	&	39.08	&	-96.58	&	366	&	Humid Grassland	&	EURODECO	&	39.33	&	16.45	&	1210	&	Mediterranean	\\
LIDET	&	10.00	&	-83.00	&		&	Wet Tropical	&	EURODECO	&	58.55	&	15.85	&	58	&	Boreal	\\
LIDET	&	18.32	&	-65.82	&	350	&	Wet Tropical	&	EURODECO	&	52.47	&	13.23	&	35	&	Temperate	\\
LIDET	&	40.28	&	-105.65	&	3160	&	Alpine Forest	&	EURODECO	&	59.73	&	14.55	&	220	&	Boreal	\\
LIDET	&	10.30	&	-84.80	&	1550	&	Wet Tropical	&	EURODECO	&	66.13	&	20.88	&	58	&	Boreal	\\
LIDET	&	33.50	&	-79.22	&	2	&	Saltmarsh Wetland	&	EURODECO	&	60.27	&	16.08	&	185	&	Boreal	\\
LIDET	&	46.00	&	-89.67	&	500	&	Temperate Coniferous	&	EURODECO	&	60.82	&	16.50	&	185	&	Boreal	\\
LIDET	&	40.05	&	-105.60	&	3650	&	Alpine Grassland	&	EURODECO	&	60.55	&	13.73	&	375	&	Boreal	\\
LIDET	&	47.83	&	-123.88	&	150	&	Temperate Coniferous	&	EURODECO	&	69.75	&	27.02	&	90	&	Boreal	\\
LIDET	&	34.33	&	-106.67	&	1572	&	Shrubland	&	EURODECO	&	59.52	&	17.27	&	30	&	Boreal	\\
LIDET	&	33.50	&	-117.75	&	500	&	Shrubland	&	EURODECO	&	50.57	&	5.98	&	370	&	Temperate	\\
LIDET	&	29.75	&	-82.50	&	35	&	Temperate Coniferous	&	EURODECO	&	65.78	&	20.62	&	135	&	Boreal	\\
LIDET	&	37.50	&	-75.67	&	0	&	Saltmarsh Wetland	&	EURODECO	&	57.42	&	15.67	&	105	&	Boreal	\\
CIDET	&	52.72	&	-106.12	&	472	&	Cool Temperate Steppe	&	EURODECO	&	60.23	&	17.47	&	580	&	Boreal	\\
CIDET	&	49.53	&	-57.83	&	50	&	Cool Temperate Subalpine Moist/Wet Forest	&	EURODECO	&	56.60	&	13.25	&	135	&	Boreal	\\
CIDET	&	47.63	&	-83.23	&	460	&	Cool Temperate Subalpine Moist/Wet Forest	&	EURODECO	&	52.33	&	22.98	&	142	&	Temperate	\\
CIDET	&	48.92	&	-54.57	&	115	&	Cool Temperate Subalpine Moist/Wet Forest	&	EURODECO	&	41.10	&	14.60	&	1100	&	Mediterranean	\\
CIDET	&	56.32	&	-94.85	&	140	&	Boreal Moist/Wet Forest	&	EURODECO	&	57.20	&	12.58	&	155	&	Boreal	\\
CIDET	&	56.32	&	-94.85	&	125	&	Boreal Moist/Wet Forest	&	EURODECO	&	64.35	&	19.77	&	260	&	Boreal	\\
CIDET	&	50.55	&	-118.83	&	650	&	Cool Temperate Moist Forest	&	EURODECO	&	60.58	&	13.57	&	435	&	Boreal	\\
CIDET	&	68.32	&	-133.53	&	73	&	Boreal Moist Forest	&	EURODECO	&	50.52	&	20.63	&	191	&	Temperate	\\
CIDET	&	51.00	&	-115.00	&	1530	&	Warm Temperate Subalpine Wet Forest	&	EURODECO	&	58.40	&	13.65	&	128	&	Boreal	\\
CIDET	&	45.42	&	-73.95	&	48	&	Cool Temperate Moist Forest	&	EURODECO	&	52.57	&	5.78	&	-5	&	Temperate	\\
CIDET	&	47.32	&	-71.13	&	670	&	Cool Temperate Subalpine Rainforest	&	EURODECO	&	58.07	&	14.13	&	245	&	Boreal	\\
CIDET	&	55.92	&	-98.62	&	288	&	Cool Temperate Subalpine Moist/Wet Forest	&	EURODECO	&	32.82	&	21.85	&	600	&	Tropical Dry	\\
CIDET	&	55.92	&	-98.62	&	260	&	Cool Temperate Subalpine Moist/Wet Forest	&	EURODECO	&	32.82	&	21.85	&	300	&	Tropical Dry	\\
CIDET	&	53.22	&	-105.97	&	476	&	Cool Temperate Moist Forest	&	EURODECO	&	66.53	&	20.18	&	280	&	Boreal	\\
CIDET	&	45.92	&	-77.58	&	173	&	Cool Temperate Moist Forest	&	EURODECO	&	60.92	&	14.02	&	350	&	Boreal	\\
CIDET	&	50.60	&	-127.33	&	100	&	Cool Temperate Wet Forest	&	EURODECO	&	40.82	&	14.48	&	250	&	Mediterranean	\\
CIDET	&	54.87	&	-66.65	&	500	&	Cool Temperate Subalpine Rain Tundra/Wet Forest	&	EURODECO	&	59.82	&	16.55	&	63	&	Boreal	\\
CIDET	&	48.63	&	-123.70	&	355	&	Cool Temperate Wet Forest	&	EURODECO	&	56.40	&	13.08	&	80	&	Boreal	\\
CIDET	&	51.83	&	-104.92	&	536.5	&	Cool Temperate Steppe	&	EURODECO	&	58.10	&	13.28	&	135	&	Boreal	\\
CIDET	&	54.60	&	-126.30	&	1100	&	Cool Temperate Subalpine Moist Forest	&	EURODECO	&	63.22	&	14.47	&	325	&	Boreal	\\
CIDET	&	60.85	&	-135.20	&	667	&	Cool Temperate Subalpine Moist Forest	&	EURODECO	&	55.65	&	13.32	&	46	&	Boreal	\\
EURODECO	&	42.73	&	-8.75	&	530	&	Mediterranean	&		&		&		&		&		\\
\hline \hline
\end{tabular}}
\end{table}

\begin{table}
\caption{The properties of litter bag datasets used to calculate the Yasso07 model solution: litter species, numbers of measurements ($N_{m}$) measurement timelines ($\Delta t$), ranges of mean annual temperature ($T_{m, min}, T_{m, max}$), amplitude of annual temperature variations ($T_{a, min}, T_{a, max}$) and precipitation ($P_{a, min}, P_{a, max}$), and initial nitrogen contents (N-cont.). Data sets with chemical composition measured as a function of time are denoted by $\star$.\label{data1}}
\resizebox{\textwidth}{!}{%
\begin{tabular}{llcccccccccl}
  \hline \hline
    Name & Species & $N_{m}$ & $\Delta t$ & $T_{m,min}$ & $T_{m,max}$ & $T_{a,min}$ & $T_{a,max}$ & $P_{a,min}$ & $P_{a,max}$ & N-cont. & Reference \\
    & & & [a] & [$^{\circ}C$] & [$^{\circ}C$] & [$^{\circ}C$] & [$^{\circ}C$] & [m] & [m] & [\%] & \\
  \hline
spine$^{\star}$	&	\textit{Pinus	sylvestris}	&	1196	&	5.42	&	3.8	&	3.8	&	11.4	&	11.4	&	0.722	&	0.722	&	0.40	&	  Berg et al., 1991a, b \\
lpine$^{\star}$	&	\textit{Pinus	contorta}	&	128	&	3.97	&	3.8	&	3.8	&	11.4	&	11.4	&	0.722	&	0.722	&	0.40	&	  Berg et al., 1991a, b \\
birch$^{\star}$	&	\textit{Betula	pubescens}	&	148	&	4.00	&	3.8	&	3.8	&	11.4	&	11.4	&	0.722	&	0.722	&	0.70	&	  Berg et al., 1991a, b \\
harad$^{\star}$	&	\textit{Pinus	sylvestris}	&	256	&	4.00	&	1.3	&	1.3	&	12.2	&	12.2	&	0.650	&	0.650	&	0.40	&	  Berg et al., 1991a, b \\
manja$^{\star}$	&	\textit{Pinus	sylvestris}	&	240	&	3.00	&	1.0	&	1.0	&	13.1	&	13.3	&	0.700	&	0.700	&	0.40	&	  Berg et al., 1991a, b \\
norrl$^{\star}$	&	\textit{Pinus	sylvestris}	&	176	&	3.33	&	1.2	&	1.2	&	11.8	&	11.8	&	0.595	&	0.595	&	0.40	&	  Berg et al., 1991a, b \\
nenne$^{\star}$	&	\textit{Pinus	sylvestris}	&	236	&	3.06	&	6.2	&	6.2	&	9.8	&	9.8	&	0.930	&	0.930	&	0.40	&	  Berg et al., 1991a, b \\
acsal	&	\textit{Acer	saccharum}	&	695	&	10.05	&	-7.0	&	26.0	&	0.8	&	20.7	&	0.209	&	3.914	&	0.81	&	  Gholz et al., 2000 \\
ambrl	&	\textit{Ammophila	breviligulata}	&	21	&	10.22	&	-3.6	&	22.1	&	2.0	&	20.7	&	0.260	&	3.500	&	0.67	&	  Gholz et al., 2000 \\
angel	&	\textit{Andropogon	gerardii}	&	35	&	9.75	&	-7.0	&	26.0	&	2.1	&	15.8	&	0.284	&	1.847	&	0.62	&	  Gholz et al., 2000 \\
aspnl	&	\textit{Populus	tremuloides}	&	126	&	6.09	&	-9.8	&	9.3	&	5.3	&	20.5	&	0.261	&	1.783	&	0.61	&	  Trofymow, 1998 \\
beecl	&	\textit{Fagus	grandifolia}	&	125	&	6.09	&	-9.8	&	9.3	&	5.3	&	20.5	&	0.261	&	1.783	&	0.66	&	  Trofymow, 1998 \\
belul	&	\textit{Betula	lutea}	&	24	&	8.98	&	-3.7	&	26.0	&	2.0	&	17.4	&	0.700	&	3.500	&	1.60	&	  Gholz et al., 2000 \\
boerl	&	\textit{Bouteloua	eriopoda}	&	35	&	8.04	&	-7.0	&	25.6	&	0.9	&	20.7	&	0.233	&	3.914	&	0.86	&	  Gholz et al., 2000 \\
bogrl	&	\textit{Bouteloua	gracilis}	&	35	&	8.01	&	-7.0	&	25.6	&	0.8	&	17.4	&	0.209	&	2.952	&	0.96	&	  Gholz et al., 2000 \\
bsprl	&	\textit{Picea	mariana}	&	126	&	6.09	&	-9.8	&	9.3	&	5.3	&	20.5	&	0.261	&	1.783	&	0.70	&	  Trofymow, 1998 \\
cedal	&	\textit{Thuja	plicata}	&	126	&	6.09	&	-9.8	&	9.3	&	5.3	&	20.5	&	0.261	&	1.783	&	0.59	&	  Trofymow, 1998 \\
cegrl	&	\textit{Ceanothus	greggii}	&	32	&	9.00	&	6.8	&	26.0	&	0.8	&	13.8	&	0.700	&	2.952	&	1.33	&	  Gholz et al., 2000 \\
conul	&	\textit{Cornus	nuttalii}	&	20	&	9.75	&	5.5	&	25.6	&	0.9	&	13.8	&	0.310	&	3.914	&	0.81	&	  Gholz et al., 2000 \\
dfirl	&	\textit{Pseudotsuga	menziesii}	&	126	&	6.09	&	-9.8	&	9.3	&	5.3	&	20.5	&	0.261	&	1.783	&	0.65	&	  Trofymow, 1998 \\
drgll	&	\textit{Drypetes	glauca}	&	623	&	10.05	&	-7.0	&	26.0	&	0.8	&	20.7	&	0.209	&	3.914	&	1.97	&	  Gholz et al., 2000 \\
fagrl	&	\textit{Fagus	grandifolia}	&	43	&	7.99	&	-7.0	&	25.6	&	0.8	&	17.4	&	0.284	&	2.952	&	0.85	&	  Gholz et al., 2000 \\
fernl	&	\textit{Pteridium	aquilinum}	&	126	&	6.09	&	-9.8	&	9.3	&	5.3	&	20.5	&	0.261	&	1.783	&	0.82	&	  Trofymow, 1998 \\
fescl	&	\textit{Festuca	hallii}	&	126	&	6.09	&	-9.8	&	9.3	&	5.3	&	20.5	&	0.261	&	1.783	&	0.64	&	  Trofymow, 1998 \\
jpinl	&	\textit{Pinus	banksiana}	&	126	&	6.09	&	-9.8	&	9.3	&	5.3	&	20.5	&	0.261	&	1.783	&	1.25	&	  Trofymow, 1998 \\
komyl	&	\textit{Kobresia	myosuroides}	&	20	&	10.04	&	-3.6	&	25.6	&	0.9	&	20.7	&	0.233	&	3.914	&	1.07	&	  Gholz et al., 2000 \\
latrl	&	\textit{Larrea	tridentata}	&	42	&	7.07	&	-7.0	&	26.0	&	0.8	&	15.5	&	0.233	&	3.500	&	2.14	&	  Gholz et al., 2000 \\
litul	&	\textit{Liriodendron	tulipifera}	&	25	&	8.03	&	-3.6	&	22.1	&	2.0	&	20.7	&	0.260	&	3.500	&	0.72	&	  Gholz et al., 2000 \\
lpilb	&	\textit{Pinus	contorta}	&	623	&	4.00	&	0.6	&	9.8	&	7.7	&	15.1	&	0.469	&	0.911	&	0.40	&	  Berg et al., 1993 \\
piell	&	\textit{Pinus	elliottii}	&	187	&	10.00	&	-7.0	&	26.0	&	0.8	&	20.7	&	0.260	&	3.914	&	0.36	&	  Gholz et al., 2000 \\
pirel	&	\textit{Pinus	resinosa}	&	536	&	10.22	&	-7.0	&	25.6	&	0.9	&	20.7	&	0.209	&	3.914	&	0.59	&	  Gholz et al., 2000 \\
pistl	&	\textit{Pinus	strobus}	&	25	&	10.05	&	-7.0	&	12.7	&	9.0	&	15.5	&	0.284	&	2.291	&	0.62	&	  Gholz et al., 2000 \\
psmel	&	\textit{Pseudotsuga	menzesii}	&	32	&	7.99	&	-7.0	&	18.0	&	9.0	&	15.5	&	0.284	&	2.291	&	0.82	&	  Gholz et al., 2000 \\
quprl	&	\textit{Quercus	prinus}	&	685	&	10.22	&	-7.0	&	26.0	&	0.8	&	20.7	&	0.209	&	3.914	&	1.03	&	  Gholz et al., 2000 \\
rhmal	&	\textit{Rhododendron	macrophyllum}	&	114	&	9.03	&	-7.0	&	25.6	&	0.8	&	15.8	&	0.233	&	3.914	&	0.42	&	  Gholz et al., 2000 \\
spall	&	\textit{Spartina	alterniflora}	&	61	&	9.04	&	-7.0	&	26.0	&	0.8	&	20.7	&	0.260	&	3.914	&	0.71	&	  Gholz et al., 2000 \\
spilb	&	\textit{Pinus	sylvestris}	&	672	&	5.39	&	-1.7	&	16.7	&	4.8	&	15.1	&	0.443	&	1.500	&	0.40	&	  Berg et al., 1993 \\
tamml	&	\textit{Larix	laricina}	&	126	&	6.09	&	-9.8	&	9.3	&	5.3	&	20.5	&	0.261	&	1.783	&	0.56	&	  Trofymow, 1998 \\
thpll	&	\textit{Thuja	plicata}	&	675	&	10.22	&	-7.0	&	26.0	&	0.8	&	20.7	&	0.209	&	3.914	&	0.62	&	  Gholz et al., 2000 \\
trael	&	\textit{Triticum	aestivum}	&	706	&	10.22	&	-7.0	&	26.0	&	0.8	&	20.7	&	0.209	&	3.914	&	0.38	&	  Gholz et al., 2000 \\
wbirl	&	\textit{Betula	papyrifera}	&	126	&	6.09	&	-9.8	&	9.3	&	5.3	&	20.5	&	0.261	&	1.783	&	0.70	&	  Trofymow, 1998 \\
  \hline \hline
\end{tabular}}
\end{table}

\end{linenumbers}


\begin{thebibliography}{100}
\bibitem[\protect\astroncite{Adair et al.}{2008}]{adair2008} Adair, E. C., Parton, W. J., Del Grosso, S. J., Silver, W. L., Harmon, M. E., Hall, S. A., Burke, I. D. and Hart, S. C. 2008. Simple three-pool model accurately describes patterns of long-term litter decomposition in diverse climates. Global Change Biology, 14, 2636-2660.
\bibitem[\protect\astroncite{Aerts}{1997}]{aerts1997} Aerts, R. 1997. Climate, leaf litter chemistry and leaf litter decomposition in terrestrial ecosystems: a tri-angular relationship. Oikos, 79, 439-449.
\bibitem[\protect\astroncite{Berg et al.}{1982}]{berg1982} Berg, B., Hannus, K., Popoff, T., and Theander, O. 1982. Changes in organic components of litter during decomposition. Long-term decomposition in a Scots pine forest. I. Canadian Journal of Botany 60, 1310-1319.
\bibitem[\protect\astroncite{Berg et al.}{1991a}]{berg1991a} Berg, B., Booltink, H., Breymeyer, A., Ewertsson, A., Gallardo, A., Holm, B., Johansson, M.-B., Koivuoja, S., Meentemeyer, V., Nyman, P., Olofsson, J., Pettersson, A.-S., Reurslag, A., Staaf, H., Staaf, I., and Uba, L. 1991a. Data on needle litter decomposition and soil climate as well as site characteristics for some coniferous forest sites, Part I, Site characteristics. Report 41, Swedish University of Agricultural Sciences, Departnent of Ecology and Environmental Research, Uppsala.
\bibitem[\protect\astroncite{Berg et al.}{1991b}]{berg1991b} Berg, B., Booltink, H., Breymeyer, A., Ewertsson, A., Gallardo, A., Holm, B., Johansson, M.-B., Koivuoja, S., Meentemeyer, V., Nyman, P., Olofsson, J., Pettersson, A.-S., Reurslag, A., Staaf, H., Staaf, I., and Uba, L. 1991b. Data on needle litter decomposition and soil climate as well as site characteristics for some coniferous forest sites, Part II, Decomposition data. Report 42, Swedish University of Agricultural Sciences, Departnent of Ecology and Environmental Research, Uppsala.
\bibitem[\protect\astroncite{Berg et al.}{1993}]{berg1993} Berg, B., Berg, M. P., Bottner, P., Box, E., Breymeyer, A., De Anta, R. C., Couteaux, M., M\"alk\"onen, E., McClaugherty, C., Meentemeyer, V., Munoz, F., Piussi, P., Remacle, J., and De Santo, A. V. 1993. Litter mass loss in pine forests of Europe and Eastern United States: some relationships with climate and litter quality. Biogeochemistry, 20, 127-159.
\bibitem[\protect\astroncite{Bosatta and \AA{}gren}{2003}]{bosatta2003} Bosatta, E. and \AA{}gren, G. I. 2003. Exact solutions to continuous-quality equation for soil organic matter turnover. Journal of Theoretical Biology, 224, 97-105.
\bibitem[\protect\astroncite{Christensen}{1996}]{christensen1996} Christensen, B. T. 1996. Matching measurable soil organic matter fractions with conceptual pools in simulation models: revision of model structure. Powlson, D. S., Smith, P., and Smith, J. U., (eds.), Evaluation of Soil Organic Matter Models, Using Existing Long Term Datasets. Springer, Berlin, pp. 143-159.
\bibitem[\protect\astroncite{Cramer et al.}{1999}]{cramer1999} Cramer, W., Kicklichter, D. W., Bondeau, A., Moore Iii, B., Churkina, G., Nemry, B., Ruimy, A., Schloss, A. L., and ThE Participants OF. ThE. Potsdam NpP. Model Intercomparison 1999. Comparing global models of terrestrial net primary productivity (NPP): Overview and key results. Global Change Biology, 5, 1-15.
\bibitem[\protect\astroncite{Davidson and Janssens}{2006}]{davidson2006} Davidson, E. A. and Janssens, I. A. 2006. Temperature sensitivity of soil carbon decomposition and feedbacks to climate change. Nature, 440, 165-173.
\bibitem[\protect\astroncite{Elliott et al.}{1996}]{elliott1996} Elliott, E. T., Paustian, K. and Frey, S. D. 1996. Modeling the measurable or measuring the modelable: a hierarchical approach to isolating meaningful soil organic matter fractionations. Powlson, D. S., Smith, P., and Smith, J. U., (eds.), Evaluation of Soil Organic Matter Models, Using Existing Long Term Datasets, Springer, Berlin, pp. 161-179.
\bibitem[\protect\astroncite{Ellison}{2004}]{ellison2004} Ellison, A. M. 2004. Bayesian inference in Ecology. Ecology Letters, 7, 509-520.
\bibitem[\protect\astroncite{Gholz et al.}{2000}]{gholz2000} Gholz, H. L., Wedin, D. A., Smitherman, S. M., Harmon, M. E., and Parton,  W. J. 2000. Long-term dynamics of pine and hardwood litter in contrasting environments: Toward a global model of decomposition. Global Change Biology, 6, 751-765.
\bibitem[\protect\astroncite{Hastings}{1970}]{hastings1970} Hastings, W. 1970. Monte Carlo sampling method using Markov chains and their applications, Biometrika 57, 97.
\bibitem[\protect\astroncite{IPCC report}{2007}]{ipcc2007} IPCC 2007: Climate Change 2007: The Physical Science Basis. Contribution of Working Group I to the Fourth Assessment Report of the Intergovernmental Panel on Climate Change. Solomon, S., Qin, D., Manning, M., Chen, Z., Marquis, M., Averyt, K. B., Tignor, M., and Miller, H. L. (eds.). Cambridge University Press, Cambridge, United Kingdom and New York, NY, USA, 996 pp.
\bibitem[\protect\astroncite{Jenkinson}{1990}]{jenkinson1990} Jenkinson, D. S. 1990. The turnover of organic carbon and nitrogen in soil. Philosophical transactions of the Royal Society, B., 329, 361-368.
\bibitem[\protect\astroncite{Jobbagy and Jackson}{2000}]{jobbagy2000} Jobb\'agy, E. G. and Jackson, R. B. 2000. The vertical distribution of soil organic carbon and its relation to climate and vegetation. Ecological Applications, 10, 423-436.
\bibitem[\protect\astroncite{Jones et al.}{2005}]{jones2005} Jones, C., McConnell, C., Coleman, K., Cox, P., Falloon, P., Jenkinson, D., and Powlson D. 2005. Global climate change and soil carbon stocks; predictions from two contrasting models for the turnover of organic carbon in soil. Global Change Biology, 11, 154-166.
\bibitem[\protect\astroncite{Kurz and Apps}{2006}]{kurz2006} Kurz, W. A. and Apps, M. J. 2006. Developing Canada’s National Forest Carbon Monitoring, Accounting and Reporting System to meet the reporting requirements of the Kyoto Protocol, Mitigation and Adaptation Strategies for Global Change, 11, 33–43.
\bibitem[\protect\astroncite{Lichter et al.}{2005}]{lichter2005} Lichter, J., Barron, S. H., Bevacqua, C. E., Finzi, A. C., Irwing, K. F., Stemmler, E. A., and Schlesinger, W. H. 2005. Soil carbon sequestration and turnover in a pine forest after six years of atmospheric CO$_{2}$ enrichment. Ecology, 86, 1835-1847.
\bibitem[\protect\astroncite{Liski et al.}{1998}]{liski1998} Liski, J., Ilvesniemi, H., M\"akel\"a, A., and Starr, M. 1998. Model analysis of the effects of soil age, fires and harvesting on the carbon storage of boreal forest soils. European Journal of Soil Science, 49, 407-416.
\bibitem[\protect\astroncite{Liski et al.}{2003}]{liski2003} Liski, J., Nissinen, A., Erhardt, M., and Taskinen, O. 2003. Climatic effects on litter decomposition from arctic tundra to tropical rainforests. Global Change Biology, 9, 575-584.
\bibitem[\protect\astroncite{Liski et al.}{2005}]{liski2005} Liski, J., Palosuo, T., Peltoniemi, M., and Siev\"anen, R. 2005. Carbon and decomposition model Yasso for forest soils. Ecological Modelling, 189, 168-182.
\bibitem[\protect\astroncite{Lloyd and Taylor}{1994}]{lloyd1994} Lloyd, J. and Taylor, J. A. 1994. On the temperature dependence of soil respiration. Functional Ecology, 8, 315-323.
\bibitem[\protect\astroncite{Manzoni et al.}{2008}]{manzoni2008} Manzoni, S., Jackson, R. B., Trofymow, J. A., and Porporato, A. 2008. The global stoichiometry of litter nitrogen mineralization. Science, 321, 684-686.
\bibitem[\protect\astroncite{Matthews}{1997}]{matthews1997} Matthews, E. 1997. Global litter production, pools, and turnover times: Estimates from measurement data and regression models. Journal of Geophysical Research, 102, 18771-18800.
\bibitem[\protect\astroncite{Meentemeyer}{1978}]{meentemeyer1978} Meentemeyer, V. 1978. Macroclimate and lignin control of litter decomposition rates. Ecology, 59, 465-472.
\bibitem[\protect\astroncite{Metropolis et al.}{1953}]{metropolis1953} Metropolis, N., Rosenbluth, A. W., Rosenbluth, M. N., Teller, A. H., and Teller, E. 1953. Equations of state calculations by fast computing machines. Journal of Chemical Physics, 21, 1087-1092.
\bibitem[\protect\astroncite{Moorhead et al.}{1999}]{moorhead1999} Moorhead, D. L., Currie, W. S., Rastetter, E. B., Parton, W. J., and Harmon, M. E. 1999. Climate and litter quality controls on decomposition: an analysis of modeling approaches. Global Biogeochemical Cycles 13, 575-589.
\bibitem[\protect\astroncite{New et al.}{2002}]{new2002} New, M., Lister, D., Hulme, M., and Makin, I. 2002. A high-resolution data set of surface climate over global land areas. Climate Research, 21, 1-25.
\bibitem[\protect\astroncite{Palosuo et al.}{2005}]{palosuo2005} Palosuo, T., Liski, J., Trofymow, J. A., and Titus, B. 2005. Litter decomposition affected by climate and litter quality - testing the Yasso model with litterbag data from the Canadian Intersite Decomposition Experiment. Ecological Modelling 189, 183-198.
\bibitem[\protect\astroncite{Parton et al.}{1987}]{parton1987} Parton, W. J., Schimel, D. S., Cole, C. V., and Ojima, D. S. 1987. Analysis of factors controlling soil organic levels of grasslands in the Great Plains. Soil Science Society of America Journal 51, 1173-1179.
\bibitem[\protect\astroncite{Parton et al.}{2007}]{parton2007} Parton, W., Silver, W L., Burke, I. C., Grassens, L., Harmon, M. E., Currie, W. S., King, J. Y., Adair, E. C., Brandt, L. A., Hart, S. C., and Fasth, B. 2007. Global-scale similarities in nitrogen release patterns during long-term decomposition. Science 315, 361-364.
\bibitem[\protect\astroncite{Schuur}{2003}]{schuur2003} Schuur, E. A. G. 2003. Productivity and global climate revisited: The sensitivity of tropical forest growth to precipitation. Ecology, 84, 1165-1170.
\bibitem[\protect\astroncite{Sitch et al.}{2003}]{sitch2003} Sitch, S., Smith, B., Prentice, I. C., Arneth, A., Bondeau, A., Cramer, W., Kaplan, J. O., Levis, S., Lucht, W., Sykes, M. T., Thonicke, K., and Venevsky, S. 2003. Evaluation of ecosystem dynamics, pland geography and terrestrial carbon cycling in the LPJ dynamical global vegetation model. Global Change Biology, 9, 161-185.
\bibitem[\protect\astroncite{Stevenson}{1982}]{stevenson1982} Stevenson, F. J. 1982. Humus chemistry. John Wiley and Sons, New York, 443 pp.
\bibitem[\protect\astroncite{Trofymow et al.}{1998}]{trofymow1998} Trofymow, J. A. and the CIDET Working Group 1998. The Canadian Intersite Decomposition ExperimenT (CIDET): Project and site establishment report. Information report BC-X-378, Pacific Forestry Centre, Victoria, Canada.
\bibitem[\protect\astroncite{Trofymow et al.}{2002}]{trofymow2002} Trofymow, J. A., Moore, T. R., Titus, B., Prescott, C. E., Morrison, I., Siltanen, M., Smith, S., Fyles, J., Wein, R., Camir\'e, C., Duschene, L., Kozak, L., Kranabetter, M., and Visser, S. 2002. Rates of litter decomposition over 6 years in Canadian forests: influence of litter quality and climate. Canadian Journal of Forest Research 32, 789-804.
\bibitem[\protect\astroncite{Tuomi et al.}{2008}]{tuomi2008} Tuomi, M., Vanhala, P., Karhu, K., Fritze, H., and Liski, J. 2008. Heterotrophic soil respiration - Comparison of different models describing its temperature dependence. Ecological Modelling, 211, 182-190.
\bibitem[\protect\astroncite{Zhang et al.}{2007}]{zhang2007} Zhang, C. F., Meng, F.-R., Trofymow, J. A., and Arp, P. A. 2007. Modeling mass and nitrogen remaining in litterbags for Canadian forest and climate conditions. Canadian Journal of Soil Science 87, 413-432.
\end{thebibliography}
\end{document}